\documentclass[11pt]{article}
\usepackage{amssymb}
\usepackage{color}
\title{\textbf{\textbf{Para-Galilean versus Galilean Noncommutative Phase Spaces}}}
\author{Ancille Ngendakumana\footnote{ancille.ngendakumana@imsp.uac.org}\\Institut de
Math\'ematiques et des Sciences Physiques, Porto-Novo, Benin\\ and \\
Joachim Nzotungicimpaye\footnote{kimpaye@kie.ac.rw}
 \\Kigali Institute of Education, Kigali, Rwanda\\ and \\ Leonard Todjihound\'e\footnote{leonardt@imsp.uac.org}\\Institut des
Math\'ematiques et des Sciences Physiques, Porto-Novo, Benin\\}
\begin{document}
\maketitle
\date
\begin{abstract}
The present paper deals with the construction of noncommutative phase spaces as coadjoint orbits of noncentral
 extensions of Galilei and Para-Galilei groups in two-dimensional space. The noncommutativity is due to
 the presence of a dual magnetic field $B^*$ in the Galilei case and of a magnetic field $B$ in the Para-Galilei case. \textbf{}
\end{abstract}
\section{Introduction}
The idea of noncommutative spacetime has been used since Snyder's original work (\cite{snyder}) and seems to appear in different approaches
in Physics.
During the last $15$ years, noncommutative mechanics has been an important subject which attracted quite a lot of attention
(\cite{horvathy11}, \cite{horvathy}, \cite{horvathy01}, \cite{romero}, \cite {vanhecke11}, \cite{wei},....).\\
Noncommutative phase spaces are defined as spaces on which coordinates satisfy the commutation relations:
 \begin{eqnarray*}\label{poisson1}
 \{ x^i,x^j \}=G^{ij}~~,~~   \{ \pi_i,x^i \}=\delta^{i}_j~~,~~ \{ \pi_i,\pi_j \}= F_{ij}
\end{eqnarray*}
where $\delta^{i}_j$ is a unit matrix, whereas $G^{ij}$ and $F_{ij}$ are functions of positions and momenta.
  The physical dimensions of $G^{ij}$ and $F_{ij}$ are respectively $M^{-1}T$ and $MT^{-1}$, where $M$ represents a mass
while $T$ represents a time.\\
Furthermore, models associated with a given symmetry group can be \\conveniently constructed using Souriau's method:
the classical phase spaces of elementary systems correspond to coadjoint orbits of their
symmetry group (\cite{souriau}).\\

The aim of this paper is to construct phase spaces endowed with modified symplectic structures by using the coadjoint orbit method.  With the
noncentrally extended Lie algebras of the Galilei and the Para-Galilei groups in two-dimensional spaces, the maximal
coadjoint orbits obtained are shown to be models of noncommutative phase spaces.
These models have the distinctive feature that the Poisson bracket of the planar positions does not vanish in the Galilei case while the planar
momenta do not commute in the Para-Galilei case. These two kinds of noncommutative phase spaces are compared in order to give all physical
distinctions between the two cases.
 Each obtained orbit is a cylinder (whose polar coordinates are action-angle $(s,\alpha)$) times a phase space whose coordinates are
the positions-momenta. In both cases, the angle $\alpha$ is constant in time.
  Moreover, in the case of canonically deformed phase space it has been proved in (\cite {walczyk}) that for Hamiltonian function of the form
\begin{eqnarray*}\label{Hamiltonian}
 H=\frac{1}{2m}(p_1^{2}+p_2^{2})+V(x^1,x^2),~~V(x^1,x^2)=\sum_iF_ix^i,~~F_i= \mbox{const.}
\end{eqnarray*}
the corresponding Newton equation $~m \frac{d^2 x_i}{dt^2}=F_i~$ remains undeformed.
 Through the Galilei and Para-Galilei symplectic realizations, the dynamic equations of motion are respectively
\begin{eqnarray*}
\frac{d^2s}{dt^2}=\frac{N^*}{m}~~,~~\frac{d^2\vec{p}}{dt^2}=m\frac{d^2\vec{q}}{dt^2}
\end{eqnarray*}
 and
\begin{eqnarray*}
\frac{d^2s}{dt^2}=\frac{N}{m}~~,~~\frac{d^2\vec{q}}{dt^2}=C\frac{d^2\vec{I}}{dt^2}
\end{eqnarray*}
where $m$ is a mass, $C$ is a spring compliance (\cite{harrie}) while $N$ and $N^*$ are quantities defined through the magnetic field $B$ and
the dual magnetic $B^*$ such that $\frac{N}{m}$ and $\frac{N^*}{m}$ are powers. \\
 It is shown that the noncommutativity of momenta induces some \\modification of the second Newton law in the Para-Galilei group case \\.

The paper is organized as follows.
In sections two and three,\\ we construct, in a two-dimensional space, the maximal coadjoint orbit of an extended Galilei group  and
of extended Para-Galilei groups respectively.  \\In section four, we give a comparative analysis of the two orbits.

\section{Noncommutative phase space on Galilei group G(2+1)}

The Galilei group $G$ in a two-dimensional space is defined
by the multiplication law
\begin{eqnarray}\label{galileilaw}
(\theta,\vec{v},\vec{x},t)(\theta~^{\prime},\vec{v}~^{\prime},\vec{x}~^{\prime},t~^{\prime})=(\theta+\theta~^{\prime},
R(\theta)\vec{v}~^{\prime}+\vec{v},R(\theta)\vec{x}~^{\prime}+\vec{v}t~^{\prime}+\vec{x},t+t~^{\prime})
\end{eqnarray}
where $\theta$ is an angle of rotations, $\vec{v}$ is a boost
vector, $\vec{x}$ is a space translation vector and $t$ is a time
translation parameter. \\

Its Lie algebra $\cal{G}$ is then generated by
the left invariant vector fields
\begin{eqnarray*}
J=\frac{\partial}{\partial\theta}~,~\vec{K}=R(-\theta)\frac{\partial}{\partial
\vec{v}}~~,~~\vec{P}=R(-\theta)\frac{\partial}{\partial
\vec{x}}~,~H=\frac{\partial}{\partial
t}+\vec{v}.\frac{\partial}{\partial \vec{x}}
\end{eqnarray*}
satisfying the nontrivial Lie brackets
\begin{eqnarray}\label{galileialgebra}
[J,K_j]=K_i\epsilon^i_j~,~[J,P_j]=P_i\epsilon^i_j~,~[K_i,H]=P_i~;~i,j=1,2.
\end{eqnarray}
\subsection{Extended group and its maximal coadjoint orbit}
Let ${\hat{\cal{G}}}$ be the extended Galilei Lie algebra satisfying relations (\ref{galileialgebra}) and
\begin{eqnarray}\label{galileialgebraextended}
[J,F_i]=F_k\epsilon^k_i~,~[P_i,H]=F_i~,~[K_i,K_j]=\frac{1}{c^2}S\epsilon_{ij}~,~[K_i,P_j]=M\delta_{ij}~~
\end{eqnarray}
where $c$ is a constant whose dimension is a velocity, the
 extending generators $M$ and $F_i$  have $L^{-2}T$ and $L^{-1}T^{-1}$ as physical dimensions respectively \\(\cite{hamermesh},
\cite{kirillov},\cite{kostant}, \cite{nzo1} ) while $S$ is dimensionless. Note that $S$ and $M$ generate
the center $ Z({\hat{\cal{G}}})$ of ${\hat{\cal{G}}}$. \\

Let $$\hat{g}=exp(\varphi S+\xi M)exp(\eta^iF_i)exp(x^iP_i+tH)exp(v^iK_i)exp(\theta J)$$
be the general element of the connected extended Galilei group $\hat{G}$. Assume also that $\hat{g}=(\beta,\vec{\eta},g)\in \hat{G}$
 with $\beta=(\varphi,\xi)$ and $g$ the general element of the Galilei group. By using relations (\ref{galileialgebra}),
(\ref{galileialgebraextended}) and the Baker-Campbell-Hausdorff formulae, we obtain that the multiplication law of $\hat{G}$
is given by
\begin{eqnarray}
 (\beta,\vec{\eta},g)(\beta^{~\prime},\vec{\eta}^{~\prime},g^{\prime})=(\beta+\beta^{~\prime}+c(g,g^{\prime}),R(\theta)\vec{\eta}^{~\prime}+
\vec{\eta}+\vec{c}~(g,g^{\prime}),gg^{\prime})
\end{eqnarray}
with $gg^{\prime}$ given by (\ref{galileilaw}) and
where
\begin{eqnarray*}
c(g,g^{\prime})=(\frac{1}{2c^2}R(-\theta)\vec{v}\times \vec{v}^{~\prime},R(-\theta)\vec{v}.\vec{x}^{~\prime}+\frac{\vec{v}^{~2}}{2}t^{\prime})
\end{eqnarray*}
while
\begin{eqnarray*}
\vec{c}~(g,g^{\prime})=\frac{1}{2}[(\vec{x}-\vec{v}t)t^{\prime}-tR(\theta)\vec{x}^{\prime}].
\end{eqnarray*}

It follows that the adjoint action of the quotient group $Q=\hat{G}/Z({\hat{G}})$ on the Lie algebra ${\hat{\cal{G}}}$ is \\
$\delta \theta^{\prime}=\delta{\theta},~~\delta t^{\prime}=\delta t$,~~$\delta \vec{v}^{~\prime}=R(\theta)\delta \vec{v} +
\epsilon(\vec{v})\delta \theta$,\\
$\delta \varphi^{\prime}=\delta \varphi+\frac{1}{c^2} R(-\theta)\vec{v}\times \delta \vec{v}-\frac{\vec{v}~^2}{2c^2}\delta \theta$,\\
$\delta\vec{x}^{~\prime}=R(\theta)\delta \vec{x}+\vec{v}\delta t+\epsilon(\vec{x}-\vec{v}t)\delta \theta-tR(\theta)\delta \vec{v}$,\\
$\delta \xi^{~\prime}=\delta \xi+R(-\theta)\vec{v}.\delta \vec{x}-R(-\theta)\vec{x}.\delta \vec{v}+\frac{\vec{v}~^2}{2}\delta t+
\vec{v}\times \vec{x}~ \delta \theta$,\\
$\delta \vec{\eta}^{~\prime}=R(\theta)\delta \vec{\eta}-t R(\theta)\delta \vec{x}+(\vec{x}-\vec{v}t)\delta t+\epsilon(\vec{\eta}-
\frac{\vec{x}t}{2}+\frac{\vec{v}t^2}{2})\delta \theta+\frac{t^2}{2}R(\theta)\delta \vec{v}$\\
with
\begin{eqnarray}\label{epsilon}
\epsilon(\vec{v})=\left(
\begin{array}{c}
v^{2}\\-v^{1}
\end{array}
\right)
\end{eqnarray}
If the duality between the extended Lie algebra ${\hat{\cal{G}}}$ and its dual ${\hat{\cal{G}}}^*$ is defined by the action
$j\delta \theta+\vec{k}.\delta \vec{v}+\vec{p}.\delta \vec{x}+E\delta t+\vec{f}.\delta {\vec{\eta}}+m\delta \xi+h\delta \varphi$
then the coadjoint action of the quotient group $Q$ on ${\hat{\cal{G}}}^*$ is
\begin{eqnarray*}
(h^{\prime},m^{\prime},\vec{f}^{~\prime},j^{\prime},\vec{k}^{~\prime},\vec{p}^{~\prime},E^{\prime})=
Ad^*_{(\vec{\eta},g)}(h,m,\vec{f},j,\vec{k},\vec{p},E)
\end{eqnarray*}
 with
\begin{eqnarray}\label{massactionforcemomentum}
m^{\prime}=m,~~h^{\prime}=h,~~\vec{f}^{~\prime}=R(\theta)\vec{f},~\vec{p}^{~\prime}=R(\theta)\vec{p}+tR(\theta)\vec{f}-m\vec{v}
\end{eqnarray}
\begin{eqnarray}\label{galileanpassage}
\vec{k}^{~\prime}=R(\theta)\vec{k}+tR(\theta)\vec{p}+\frac{t^2}{2}R(\theta)\vec{f} +m(\vec{x}-\vec{v}t)+\frac{h}{c^2}\epsilon(\vec{v})
\end{eqnarray}
\begin{eqnarray}\label{galileanenergy}
E^{\prime}=E-\vec{v}.R(\theta)\vec{p}-\vec{x}.R(\theta)\vec{f}+\frac{m\vec{v}~^2}{2}
\end{eqnarray}
\begin{eqnarray}\label{galileianangularmomentum}
j^{\prime}=j+\vec{x}\times R(\theta)\vec{p}+\vec{v}\times R(\theta)\vec{k}+{\vec{\eta}}\times R(\theta)\vec{f}+
\frac{\vec{x}t}{2}\times R(\theta)\vec{f}+m\vec{v}\times \vec{x}
\end{eqnarray}
The observables $j, \vec{k}, \vec{p},E, \vec{f}, m$  and $h$ have respectively the physical \\dimensions of an angular momentum, a static momentum,
 a linear \\momentum, an energy, a force, a mass and an action. \\

 Let the position vector $\vec{q}$ and the dual magnetic field $B^*$ (\cite{ancilla}) be defined by
 \begin{eqnarray}\label{dualmagnetic}
 \vec{q}=\frac{\vec{k}}{m},~~B^*=\frac{1}{e^*m\omega},
 \end{eqnarray}
where $e^*$ is a dual charge.\\
The maximal coadjoint orbit of $Q$ on ${\hat{\cal{G}}^*}$ denoted by ${\cal}{O}_{(m,B^*,f,U)}$ is then characterized by the two trivial
invariants $m$ and $h$ and by two nontrivial invariants: the force intensity $f$ and the energy $U$ given by
\begin{eqnarray*}
f=||\vec{f}||,~~U=E-\frac{\vec{p}~^2}{2m}+\vec{f}.\vec{q}+e^*B^*\vec{f}\times\vec{p}
\end{eqnarray*}
where the relation $h\omega=mc^2$ has been used.  In the basis\\
 $(J,F_1,K_1,P_1,K_2,P_2,F_2,H,M,S)$ of the extended Galilei Lie algebra, the restriction of the Kirillov's matrix on the orbit is
\begin{eqnarray*}
\Omega=\left (\begin{array}{cccccc}
0&f sin\alpha&mq^2&p_2&-mq^1&-p_1\\-f sin\alpha&0&0&0&0&0\\-mq^2&0&0&m&e^*m^2B^*&0\\-p_2&0&-m&0&0&0\\mq^1&0&-e^*m^2B^*&0&0&m\\p_1&0&0&0&-m&0
\end{array}
\right)
\end{eqnarray*}
where $f_1=f cos\alpha~,~f_2=f sin\alpha$.\\

Its inverse is
\begin{eqnarray*}
\Omega^{-1}=\frac{1}{f~ sin\alpha}\left (\begin{array}{cccccc}
0&-1&0&0&0&0\\1&0&-\frac{p_2}{m}&q^2-e^*B^*p_1&\frac{p_1}{m}&-q^1-e^*B^*p_2\\
0&\frac{p_2}{m}&0&-\frac{f~ sin\alpha}{m}&0&0\\0&-q^2+e^*B^*p_1&\frac{f~ sin\alpha}{m}&0&0&e^*B^*f~\sin \alpha\\
0&-\frac{p_1}{m}&0&0&0&-\frac{f~ sin\alpha}{m}\\0&q^1+e^*B^*p_2&0&-e^*B^*f~\sin\alpha&\frac{f~ sin\alpha}{m}&0
\end{array}
\right)
\end{eqnarray*}
where $\vec{A^*}=\frac{1}{2}\vec{B^*}\times \vec{p} ~$ is the dual magnetic potential (\cite{ancilla})  while
 $\vec{B^*}=B^*\vec{n}$ with $\vec{n}$ the unit vector perpendicular to the plane.\\The orbit is then equipped with the symplectic form
\begin{eqnarray}\label{symplecticformnc1}
\sigma_1=\sigma_0+G^{ij}dp_i\wedge dp_j
\end{eqnarray}
where  $\sigma_0=ds\wedge d\alpha+dp_i\wedge dq^i$, $G^{ij}=e^*B^*\epsilon^{ij}$ and $s=j-\vec{p}\times (\vec{q}-e^*\vec{A}^*)$ is
the sum of the orbital momentum $\vec{l}=\vec{q}\times \vec{p}$, the angular momentum $j$ and an extra term $\vec{p}\times e^*\vec{A}^*$
associated to the dual magnetic field $B^*$ (\cite{horvathy11}, \cite{horvathy}).\\
The Poisson brackets (of two functions $H$ and $f$ defined on the orbit)\\
corresponding to the symplectic structure (\ref{symplecticformnc1}), are given by
\begin{eqnarray*}
\{H,f\}=\frac{\partial H}{\partial s}\frac{\partial f}{\partial \alpha}-\frac{\partial H}{\partial \alpha}\frac{\partial f}{\partial s}+
\frac{\partial H}{\partial p_i}\frac{\partial f}{\partial q^i}-\frac{\partial H}{\partial q^i}\frac{\partial f}{\partial p_i}-
e^*B^*\epsilon^{ij}\frac{\partial H}{\partial q^i}\frac{\partial f}{\partial q^j}
\end{eqnarray*}
Then
\begin{eqnarray*}\label{poissongalilei}
\{s,p_j\}=p_i\epsilon^i_j~,~\{s,q^i\}=\epsilon^i_j(q^j-e^*A^{*j})~,~\{q^i,q^j\}=-e^*B^*\epsilon^{ij}~,~\{p_i,q^j\}=\delta^j_i~
\end{eqnarray*}
are the Poisson brackets within the coordinates on the orbit.  We observe that $\alpha$ commute with all the other coordinates, that
 the momenta commute and form a vector,
that positions do not commute and do not form a vector due to the presence of the dual magnetic field $B^*$.\\
Note also that $(s,p_i,\alpha,\tilde{q}~^i=q^i-e^*A^{*i})$ are canonical phase coordinates on the coadjoint orbit and
 that
\begin{eqnarray*}
\{s,A^*_j\}=A^*_i\epsilon^i_j~,~~\{q^i,A^*_j\}=\frac{B^*}{2}\epsilon^i_j.
\end{eqnarray*}
\subsection{Symplectic realization and equations of motion}
 Let  $(s^{\prime},\vec{p}^{~\prime},\alpha^{\prime},\vec{q}^{~\prime})=D_{(\vec{\eta},\theta,\vec{v},\vec{x},t)}(s,\vec{p},\alpha,\vec{q})$ be
the symplectic realization of the extended Galilei group on its coadjoint orbit.
Use of the relations (\ref{massactionforcemomentum}) to (\ref{galileianangularmomentum}) gives rise to
\begin{eqnarray*}
\alpha^{\prime}=\alpha+\theta~,~\vec{q}^{~\prime}=R(\theta)\vec{q}+\frac{1}{m}(R(\theta)\vec{p}-m\vec{v})t+
\frac{R(\theta)\vec{f}}{m}\frac{t^2}{2}+\vec{x}-e^*m\vec{v}\times \vec{B}^*
\end{eqnarray*}
\begin{eqnarray*}
s^{\prime}=s+\frac{K^*}{m}t+\frac{N^*}{m}\frac{t^2}{2}+(\vec{\eta}-\frac{\vec{x}t}{2}+\frac{\vec{v}t^2}{2})\times R(\theta)\vec{f}-\frac{e^*B^*m^2\vec{v}^2}{2}
\end{eqnarray*}
\begin{eqnarray*}
\vec{p}^{~\prime}=R(\theta)\vec{p}+R(\theta)\vec{f}t-m\vec{v}
\end{eqnarray*}
where $E^*_0=\frac{K^*}{m}$ and $P^*_o=\frac{N^*}{m}$ are respectively an energy and a power with $K^*$ and $N^*$ given by:
\begin{eqnarray}\label{energypower}
K^*=m\vec{q}\times \vec{f}+e^*mB^*\vec{p}.\vec{f}~,~N^*=\vec{f}\times \vec{p}-e^*mB^*\frac{\vec{f}~^2}{2}
\end{eqnarray}
It follows that the evolution with respect to the time $t$ is given by
\begin{eqnarray}\label{configurationevolution1}
\alpha(t)=\alpha,~\vec{q}~(t)=\vec{q}+\frac{\vec{p}}{m}t+\frac{\vec{f}}{m}\frac{t^2}{2}
\end{eqnarray}
and
\begin{eqnarray}\label{momentaevolution1}
s(t)=s+\frac{K^*}{m}t+\frac{N^*}{m}\frac{t^2}{2},
~\vec{p}~(t)=\vec{p}+\vec{f}t
\end{eqnarray}
The corresponding Hamiltonian vector field is
\begin{eqnarray*}
X_H=\frac{\vec{p}}{m}.\frac{\partial}{\partial \vec{q}}+\vec{f}.\frac{\partial}{\partial \vec{p}}+\frac{K^*}{m}\frac{\partial}{\partial s}
\end{eqnarray*}
 and then the Hamiltonian function, given by
\begin{eqnarray*}
H=\frac{\vec{p}^{~2}}{2m}+V(\vec{q},\alpha)+e^*\vec{p}.(\vec{f}\times \vec{B}^*),
\end{eqnarray*}
is the sum of a kinetic energy $T(\vec{p})=\frac{\vec{p}^2}{2m}$, of a potential energy \\$V(\vec{q},\alpha)=-\vec{f}.\vec{q}-
\frac{K^*}{m}\alpha$ depending on the position-angle variables $(\vec{q},\alpha)$ and of an exotic energy $E^*_{exotic}=e^*\vec{p}.
(\vec{f}\times \vec{B}^*)$ depending on the dual magnetic field.\\
The equations of motion are then given by
\begin{eqnarray}\label{forcemomentum}
\left(
\begin{array}{c}
\vec{p}(t)\\K^*(t)
\end{array}
\right)=m\frac{d}{dt}\left(
\begin{array}{c}
\vec{q}(t)\\ s(t)
\end{array}
\right)~,~
\frac{d}{dt}\left(
\begin{array}{c}
\vec{p}(t)\\ \alpha(t)
\end{array}
\right)= \left(
\begin{array}{c}
\vec{f}\\0
\end{array}
\right)
\end{eqnarray}
The equations (\ref{forcemomentum}) define the linear momentum  $\vec{p}~(t)=m\frac{d\vec{q}~(t)}{dt}$ , the force
$\vec{f}=\frac{d\vec{p}}{dt}$ and the quantity
$K^*(t)=m\frac{ds}{dt}$ whose dimension is that of a linear momentum squared.
Moreover
\begin{eqnarray}\label{accelleration}
\frac{d^2}{dt^2}\left(
\begin{array}{c}
\vec{p}\\\alpha
\end{array}
\right)=\left(
\begin{array}{c}
\vec{0}\\0
\end{array}
\right)
,
\frac{d^2}{dt^2}\left(
\begin{array}{c}
\vec{q}\\s
\end{array}
\right)=\frac{1}{m}\left(
\begin{array}{c}
\vec{f}\\N^*
\end{array}
\right)
\end{eqnarray}
We interpret the equations (\ref{accelleration}) as the Euler-Newton's second law (\cite{wei}) \\associated to the above noncentral extended Galilei
group.\\
With the coadjoint orbit method applied to this group, the phase space \\ obtained is a six-dimensional
phase space where one coordinates commute with the others and where the positions do not commute due to the \\noncommutativity of the generators
of pure Galilei transformations.\\ This noncommutativity is measured by the presence of the dual magnetic filed $B^*$ through the Poisson brackets
 (\ref{poissongalilei}). With the Galilei group, same results have been obtained in (\cite{horvathy}) but by using the two centrally extended
 Lie algebra of this group. \\
In the following section, we study the case of the Para-Galilei groups.\\ We show that the noncommutativity of the
 momenta induces some modification of the Newton's second law
(\cite{horvathy}, \cite{romero},\cite{wei}).

 \section{Noncommutative phase spaces on the Para-Galilei groups $G^{\prime}_{\pm}(2+1)$}

The non-relativistic Para-Galilei groups are obtained through a spacetime \\contraction (\cite{5bacry},\cite{inonu}) of the Newton-Hooke groups
or through a space-velocity contraction of the Para-Poincar\'e group.  They contract themselves by a velocity-time contraction in the Static group.\\

The Para-Galilei groups $G^{\prime}_{\pm}$ in two-dimensional spaces are defined
by the multiplication law
\begin{eqnarray}\label{paragalileilaw}
(\theta,\vec{v},\vec{x},t)(\theta^{~\prime},\vec{v}^{~\prime},\vec{x}^{~\prime},t^{~\prime})=(\theta+\theta^{~\prime},
R(\theta)\vec{v}^{~\prime}+\vec{v}\pm\omega^2 \vec {x}t^{~\prime},R(\theta)\vec{x}^{~\prime}+\vec{x},t+t^{~\prime})
\end{eqnarray}
where $\theta$ is an angle of rotations, $\vec{v}$ is a boost
vector, $\vec{x}$ is a space translation vector and $t$ is a time
translation parameter.\\

 Their Lie algebras $\cal{G}^{\prime}_{\pm}$ are then generated by
the left invariant vector fields
\begin{eqnarray*}
J=\frac{\partial}{\partial
\theta}~,~\vec{K}=R(-\theta)\frac{\partial}{\partial
\vec{v}}~~,~~\vec{P}=R(-\theta)\frac{\partial}{\partial
\vec{x}}~~~~H=\frac{\partial}{\partial
t}\pm\omega^2 \vec{x}\frac{\partial}{\partial
\vec{v}}~~
\end{eqnarray*}
satisfying the nontrivial Lie brackets
\begin{eqnarray}\label{para-galiliealgebra}
[J,K_j]=K_i\epsilon^i_j~,~[J,P_j]=P_i\epsilon^i_j~,~[P_i,H]=\pm\omega^2 K_i~
\end{eqnarray}
\subsection{Extended groups and their maximal coadjoint orbits}
 The extended Para-Galilei Lie algebras ${\hat{\cal{G}}^{\prime}}_{\pm}$ satisfy the nontrivial Lie brackets
(\ref{para-galiliealgebra}) and
\begin{eqnarray}\label{extendedparagalileialgebra}
[J,\Pi_i]=\Pi_k\epsilon^k_i~,~[K_i,H]=\Pi_i~,~[P_i,P_j]=\frac{1}{r^2}S\epsilon_{ij}~,~[K_i,P_j]=M\delta_{ij}~~
\end{eqnarray}
where $r$ is a constant whose dimension is a length (i.e the radius of the universe).  Note that $S$ and $M$ generate the center
of ${\hat{\cal{G}}^{\prime}}_{\pm}$.\\

 Let $$\hat{g}=exp(\varphi S+\xi M)exp({\cal}{l}^i\Pi_i)exp(v^iK_i+tH)exp(x^iP_i)exp(\theta J)$$
be the general element of the connected extended Para-Galilei groups ${\hat{G}}^{\prime}_{\pm}$.
As in the case of the Galilei group, the relations (\ref{para-galiliealgebra}) , (\ref{extendedparagalileialgebra}) and
the Baker-Campbell-Hausdorff
formulae give rise to the multiplication law
\begin{eqnarray}
(\beta,{\cal}{\vec{l}},g)(\beta^{\prime},{\cal}{\vec{l}}^{~\prime},g^{\prime})=(\beta+\beta^{\prime}+c(g,g^{\prime}),
{\cal}{\vec{l}}+R(\theta){\cal}{\vec{l}}^{~\prime}+\vec{c}~(g,g^{\prime}),gg^{\prime})
\end{eqnarray}
with $gg^{\prime}$ given by (\ref{paragalileilaw}) and
where
\begin{eqnarray*}
c(g,g^{\prime})=(\frac{1}{2r^2}R(-\theta)\vec{x}\times \vec{x}^{~\prime},-R(-\theta)\vec{x}.\vec{v}^{~\prime}\mp\frac{\omega^2\vec{x}~^2}{2}
t^{\prime})
\end{eqnarray*}
while
\begin{eqnarray*}
\vec{c}~(g,g^{\prime})=\frac{1}{2}[(\vec{v}\mp \omega^2\vec{x}t)t^{\prime}-tR(\theta)\vec{v}^{~\prime}].
\end{eqnarray*}

It follows that the adjoint action of the quotient groups $Q^{\prime}_{\pm}=\hat{G}^{\prime}_{\pm}/Z({\hat{G}}^{\prime}_{\pm})$ on the extended
 Para-Galilei Lie algebras
${\hat{\cal{G}}^{\prime}}_{\pm}$ is \begin{eqnarray*}
\delta \theta^{\prime}=\delta{\theta}~,~\delta t^{\prime}=\delta t~,~\delta\vec{x}^{~\prime}=R(\theta)\delta \vec{x}+
\epsilon(\vec{x})\delta \theta
\end{eqnarray*}
\begin{eqnarray*}
\delta \vec{v}^{~\prime}=R(\theta)\delta \vec{v}\pm \omega^2\vec{x}\delta t+
\epsilon(\vec{v}\mp \omega^2\vec{x}t)\delta \theta \mp \omega^2 t R(\theta)\delta \vec{x}
\end{eqnarray*}
\begin{eqnarray*}
\delta {\cal}{\vec{l}}^{~\prime}=R(\theta){\cal}{\vec{l}}-t R(\theta)\delta \vec{v}\pm \frac{\omega^2t^2}{2}R(\theta)\delta \vec{x}+
(\vec{v}\mp\omega^2\vec{x}t)\delta t+\epsilon({\cal}{\vec{l}}-\frac{t}{2}(\vec{v}\mp \omega^2\vec{x}t))\delta \theta
\end{eqnarray*}
\begin{eqnarray*}
\delta \varphi^{\prime}=\delta \varphi+\frac{1}{r^2} R(-\theta)\vec{x}\times \delta \vec{x}-\frac{\vec{x}~^2}{2r^2}\delta \theta~\\
~\delta \xi^{\prime}=\delta \xi-R(-\theta)\vec{x}.\delta \vec{v}+R(-\theta)\vec{v}.\delta \vec{x}\mp\frac{\omega^2\vec{x}^2}{2}\delta t+
\vec{v}\times \vec{x}~ \delta \theta
\end{eqnarray*}
where $\epsilon(\vec{v})$ is given by (\ref{epsilon}).
If the duality between the extended Lie algebras ${\hat{\cal{G}}^{\prime}}_{\pm}$ and their duals ${\hat{\cal{G}}^{*{\prime}}}_{\pm}$ gives rise
to the action \\$j\delta \theta+\vec{k}.\delta \vec{v}+
\vec{I}.\delta \vec{x}+E\delta t+\vec{p}.\delta {\cal}{\vec{l}}+m\delta \xi+h\delta \varphi$, then the coadjoint actions are such that
\begin{eqnarray}\label{massaction}
m^{\prime}=m~,~h^{\prime}=h
\end{eqnarray}
\begin{eqnarray}\label{passage}
\vec{p}^{~\prime}=R(\theta)\vec{p}~,~~\vec{k}^{~\prime}=R(\theta)\vec{k}+tR(\theta)\vec{p}+m\vec{x}
\end{eqnarray}
\begin{eqnarray}\label{linearmomentum}
\vec{I}^{~\prime}=R(\theta)\vec{I}\pm\omega^2tR(\theta)\vec{k}\pm\frac{\omega^2t^2}{2}R(\theta)\vec{p} +
eB\epsilon(\vec{x})-m(\vec{v}\mp\omega^2\vec{x}t)
\end{eqnarray}
\begin{eqnarray}\label{energy}
E^{\prime}=E\mp \omega^2\vec{x}.R(\theta)\vec{k}-\vec{v}.R(\theta)\vec{p}\mp\frac{m\omega^2\vec{x}~^2}{2}
\end{eqnarray}
\begin{eqnarray}\label{angularmomentum}
j^{~\prime}=j+\vec{x}\times R(\theta)\vec{I}+\vec{v}\times R(\theta)\vec{k}+{\cal}{\vec{l}}\times R(\theta)\vec{p}+
\frac{\vec{v}t}{2}\times R(\theta)\vec{p}+m\vec{v}\times \vec{x}
\end{eqnarray}
Define the vector $\vec{q}$ by (\ref{dualmagnetic}) and the magnetic field $B$ by
\begin{eqnarray}
B=\frac{m\omega}{e}
\end{eqnarray}
where $e$ is the electric charge. The coadjoint orbits denoted by ${\cal}{O}_{(m,B,p,U_{\pm})}$ are characterized by two
trivial invariants $m$ and
$B$ and by two nontrivial invariants $p$ and $U_{\pm}$ respectively given by:
\begin{eqnarray*}
p=||\vec{p}||,~~U_{\pm}=E\pm\frac{m\omega^2\vec{q}~^2}{2}-\vec{p}.\frac{\vec{I}}{m
}+eB\vec{p}\times\vec{q}
\end{eqnarray*}
where we have used the relation $h\omega=mc^2$.
 In the basis \\$(J,P_1,K_1,\Pi_1,K_2,\Pi_2,P_2,H,M,S)$ of the extended Para-Galilei Lie algebras, the restriction of the Kirillov's form on the
coadjoint orbit is then
\begin{eqnarray*}
\Omega=\left (\begin{array}{cccccc}
0&p~ sin\alpha&mq^2&I_2&-mq^1&-I_1\\-p~ sin\alpha&0&0&0&0&0\\-mq^2&0&0&m&0&0\\-p_2&0&-m&0&0&eB\\mq^1&0&0&0&0&m\\I_1&0&0&-eB&-m&0
\end{array}
\right)
\end{eqnarray*}
where $p_1=p~ cos\alpha~,~p_2=p~ sin\alpha$. The inverse of $\Omega$ is
\begin{eqnarray*}
\Omega^{-1}=\frac{1}{p~ sin\alpha}\left (\begin{array}{cccccc}
0&-1&0&0&0&0\\1&0&-\frac{I_2}{m}-\frac{eBq^1}{m}&q^2&\frac{I_1}{m}-\frac{eBq^2}{m}&-q^1\\0&\frac{I_2}{m}+\frac{eBq^1}{m}&0&-
\frac{p~ sin\alpha}{m}&\frac{eBp~ sin\alpha}{m^2}&0\\0&-q^2&\frac{p ~sin\alpha}{m}&0&0&0\\0&-\frac{I_1}{m}+\frac{eBq^2}{m}&-
\frac{eBp~ sin\alpha}{m^2}&0&0&-\frac{p~ sin\alpha}{m}\\0&q^1&0&0&\frac{p~ sin\alpha}{m}&0
\end{array}
\right)
\end{eqnarray*}
We then verify that the symplectic form on the orbit ${\cal}{O}_{(m,B,p,U_{\pm})}$ is
\begin{eqnarray}\label{symplecticformnc}
\sigma_1^{\prime}=\sigma_0^{\prime}+F_{ij}dq^i\wedge dq^j
\end{eqnarray}
where $\sigma_0^{\prime}=ds\wedge d\alpha+dI_i\wedge dq^i$, $F_{ij}=eB\epsilon_{ij}$ and $s =j-\vec{q}\times (\vec{I}+e\vec{A})$ is
 the sum of the angular momentum $j$, the orbital angular momentum $\vec{l}=\vec{q}\times \vec{I}$
and an extra term $eB\frac{\vec{q}~^2}{2}$ associated to the magnetic field $B$, $\vec{A}=\frac{1}{2}\vec{B}\times \vec{q}$ being the
magnetic potential \cite{ancilla} with $\vec{B}=B\vec{n}$.\\

The Poisson brackets of two functions $H$ and $f$ on the orbit \\corresponding to the symplectic form
 (\ref{symplecticformnc}) is
\begin{eqnarray*}
\{H,f\}=\frac{\partial H}{\partial s}\frac{\partial f}{\partial \alpha}-\frac{\partial H}{\partial \alpha}\frac{\partial f}{\partial s}+
\frac{\partial H}{\partial I_i}\frac{\partial f}{\partial q^i}-\frac{\partial H}{\partial q^i}\frac{\partial f}{\partial I_i}-
eB\epsilon_{ij}\frac{\partial H}{\partial I_i}\frac{\partial f}{\partial I_j}
\end{eqnarray*}
and the non trivial Poisson brackets within the coordinates are
\begin{eqnarray}\label{possonparagalilei}
\{s,I_j\}=(I_i-eA_i)\epsilon^i_j~,~\{s,q^i\}=\epsilon^i_jq^j~,~\{I_i,I_j\}=-eB\epsilon_{ij}~,~\{I_i,q^j\}=\delta^j_i~
\end{eqnarray}
This means that $\alpha$ commute with all the other coordinates, that the positions coordinates commute and form a vector
while momenta do not commute and do not form a vector due to the presence of the magnetic field $B$.\\

Note that $(s,\tilde{I}_i=I_i-eA_i,\alpha,q^i)$ are the canonical phase coordinates on the coadjoint orbit and that
we have moreover these Poisson brackets
\begin{eqnarray}
\{s,A_j\}=-A_i\epsilon^i_j~,~\{I_i,A_j\}=\frac{B}{2}\epsilon_{ij}
\end{eqnarray}

\subsection{Symplectic realizations and equations of motion}

 Let the symplectic realizations of the extended Para-Galilei groups on their coadjoint orbits be given by
$(s^{\prime},\vec{p}^{~\prime},\alpha^{\prime},\vec{q}^{~\prime})=D_{({\cal}{\vec{l}},\theta,\vec{v},\vec{x},t)}(s,\vec{p},\alpha,\vec{q})$ .
 By using relations (\ref{passage}) to (\ref{angularmomentum}), we obtain
\begin{eqnarray*}
\alpha^{\prime}=\alpha+\theta~,~\vec{q}^{~\prime}=R(\theta)\vec{q}+\frac{R(\theta)\vec{p}}{m}t+\vec{x}
\end{eqnarray*}
\begin{eqnarray*}
s^{\prime}=s+\frac{K}{m} t+\frac{N}{m} \frac{t^2}{2}+(\vec{{\cal}{l}}-\frac{\vec{v}t}{2}\pm\frac{\omega^2t^2\vec{x}}{2})\times R(\theta)\vec{p}+
eB\frac{{\vec{x}~^2}}{2}
\end{eqnarray*}
\begin{eqnarray*}
\vec{I}^{~\prime}=R(\theta)\vec{I}\pm m\omega^2[R(\theta)\vec{q}+\vec{x}]t\pm \omega^2R(\theta)\vec{p}~\frac{t^2}{2} +eB\epsilon(\vec{x})-m\vec{v}
\end{eqnarray*}
where $E_0=\frac{K}{m}$ and $P_0=\frac{N}{m}$ are respectively an energy and a power and where $K$ and $N$ are given by
\begin{eqnarray}
K=\vec{I}\times \vec{p}-eB\vec{p}.\vec{q}~,~N=\pm m\omega^2\vec{q}\times \vec{p}\mp eB\frac{\vec{p}^2}{2m}
\end{eqnarray}
 The evolution with respect to the time  $t$ is
\begin{eqnarray}\label{configurationevolution2}
\alpha(t)=\alpha~,~\vec{q}~(t)=\vec{q}+\frac{\vec{p}}{m}t
\end{eqnarray}
and
\begin{eqnarray}\label{momentaevolution2}
s(t)=s+\frac{K}{m}t+\frac{N}{m}\frac{t^2}{2} ~,~\vec{I}(t)=
\vec{I}\pm m\omega^2\vec{q}t\pm \omega^2\vec{p}~\frac{t^2}{2}
\end{eqnarray}
It follows that the corresponding Hamiltonian vector field is
\begin{eqnarray*}
X_H=\frac{\vec{p}}{m}.\frac{\partial}{\partial \vec{q}}\pm m\omega^2\vec{q}.\frac{\partial}{\partial \vec{I}}+
\frac{K}{m}\frac{\partial}{\partial s}
\end{eqnarray*}
and then the Hamiltonian function (energy) is
\begin{eqnarray*}
H=\vec{I}.\frac{\vec{p}}{m}\mp\frac{m\omega^2\vec{q}^2}{2}-\frac{K}{m}\alpha+e\vec{p}.(\vec{B}\times \vec{q})
\end{eqnarray*}
i.e a sum of a kinetic term $T(\vec{p})=\vec{I}.\frac{\vec{p}}{m}$, of a potential energy\\ $V(\vec{q},\alpha)=
\mp\frac{m\omega^2\vec{q}^2}{2}-\frac{K}{m}\alpha$~ depending on the position-angle variables $(\vec{q},\alpha)$ and of an exotic energy
$E_{exotic}=e\vec{p}.(\vec{f}\times \vec{B})$ depending on the magnetic field.\\
The equations of motion are then given by
\begin{eqnarray}\label{positionimpulse}
\frac{d}{dt}\left(
\begin{array}{c}
\vec{I}(t)\\ s(t)
\end{array}
\right)=\pm m\omega^2\left(
\begin{array}{c}
\vec{q}\\\pm \frac{K(t)}{m^2\omega^2}
\end{array}
\right)~,~
\frac{d}{dt}\left(
\begin{array}{c}
\vec{q}(t)\\ \alpha(t)
\end{array}
\right)=\frac{1}{m}\left(
\begin{array}{c}
\vec{p}\\0
\end{array}
\right)
\end{eqnarray}
The equations (\ref{positionimpulse}) define the linear momentum  $\vec{p}~(t)=m\frac{d\vec{q}~(t)}{dt}$,  the position
 $\vec{q}=C\frac{d\vec{I}}{dt}$ where $C=\pm\frac{1}{m\omega^2}$ is a spring compliance i.e the inverse of a spring constant (\cite{harrie}) and
the quantity $K(t)=m\frac{ds}{dt}$ whose dimension is that of a linear momentum squared.
Moreover,
\begin{eqnarray}\label{accellerationb}
\frac{d^2}{dt^2}\left(
\begin{array}{c}
\vec{I}\\s
\end{array}
\right)=\frac{1}{m}\left(
\begin{array}{c}
\pm m\omega^2\vec{p}\\N
\end{array}
\right)
,
\frac{d^2}{dt^2}\left(
\begin{array}{c}
\vec{q}\\ \alpha
\end{array}
\right)=\left(
\begin{array}{c}
\vec{0}\\0
\end{array}
\right)
\end{eqnarray}
We interpret the equations (\ref{accellerationb}) as the modified Euler-Newton's second law (\cite{wei}) associated to the Para-Galilei groups.\\
The phase space obtained is a six-dimensional phase space where one coordinates commute with the others and where the momenta do not
commute due to the noncommutativity of the generators of space translations in the extended Para-Galilei groups. This noncommutativity is
 measured by the presence of the magnetic field $B$ through the Poisson brackets (\ref{possonparagalilei}).
\section{Comparative analysis }
 In both cases the coadjoint orbit is a six-dimensional symplectic manifold i.e the product of
a two dimensional cylinder (parametrized by an angle $\alpha$ and a conjugate action $s$) with a four dimensional phase space parametrized
by the position-momentum coordinates. These coadjoint orbits are denoted ${\cal}{O}_{(m,f,B^*,U)}$ for the Galilei group
 and ${\cal}{O}_{(m,p,B,U_{\pm})}$ for the Para-Galilei groups where $B$
and $B^*$ are respectively a magnetic and a dual magnetic field.  The polar coordinates of the points on the basis circle of
the cylinder are $(f,\alpha)$ for the orbit ${\cal}{O}_{(m,f,B^*,U)}$  and $(p,\alpha)$  for ${\cal}{O}_{(m,p,B,U_{\pm})}$ where $f$ is a
constant force intensity while $p$ is a constant linear momentum. The orbit ${\cal}{O}_{(m,f,B^*,U)}$ is equipped with the symplectic
form $\sigma_1=ds\wedge d\alpha+dp_i\wedge dq^j+e^*B^*dp_i\wedge dp_j$ and is then a noncommutative symplectic manifold where the
positions $q^i$ do not commute while the orbit ${\cal}{O}_{(m,p,B,U_{\pm})}$  endowed with the symplectic
 form $\sigma_1^{\prime}=ds\wedge d\alpha+dI_i\wedge dq^j+eBdq^i\wedge dq^j$ is a noncommutative symplectic
 manifold where the momenta $I_i$ do not commute. Kinematically, the positions $q^i$ on ${\cal}{O}_{(m,f,B^*,U)}$ behave as the momenta $I_i$
on ${\cal}{O}_{(m,p,B,U_{\pm})}$ (\ref{configurationevolution1}) and (\ref{momentaevolution2})) whereas the momenta $p_i$ on
${\cal}{O}_{(m,f,B^*,U)}$ behave as the positions $q^i$ on ${\cal}{O}_{(m,p,B,U_{\pm})}$ (\ref{momentaevolution1}) and
(\ref{configurationevolution2})). The former are quadratic in time while the last ones are linear in time. The action $s$ conjugated to
the angle is quadratic in time in the two cases. Note that the angle $\alpha$ is constant in time in the two cases.\\The dynamic equations
for the system described by the Galilean orbit are
\begin{eqnarray}
\frac{d^2s}{dt^2}=\frac{N^*}{m}~,~\frac{d\vec{p}}{dt}=m\frac{d^2\vec{q}}{dt^2}
\end{eqnarray}
 while they are
  \begin{eqnarray}
\frac{d^2s}{dt^2}=\frac{N}{m}~,~\frac{d\vec{q}}{dt}=C\frac{d^2\vec{I}}{dt^2}
\end{eqnarray}
for the system described by the Para-Galilean orbits, where $C=\pm \frac{1}{m\omega^2}$ is the inverse of the spring constant.
 As the equation $\frac{d\vec{p}}{dt}=m\frac{d^2\vec{q}}{dt^2}$ is called the Galilei-Newton law (\cite{nzo1}) for a massive
particle with mass $m$, the
 law  $\frac{d\vec{q}}{dt}=C\frac{d^2\vec{I}}{dt^2}$ can be called the Para-Galilei-Newton law for a spring whose contant
 $C$ is the inverse of the usual Hooke contant. The following table summarizes these constructions in both cases.
\begin{eqnarray*}
\begin{tabular}{|c|c|c|}
\hline
\textbf{Group}&\textbf{Galilei~group}&\textbf{Para-Galilei~groups}\\
\hline
 Invariants(non trivial)& $f,~U=e-\frac{\vec{p}^2}{2m}+\vec{f}.\vec{q}+e^*B^*\vec{f}\times\vec{p}$& $p,~U_{\pm}=e\pm\frac{m\omega^2\vec{q}^2}{2}$\\
&& $-\vec{p}.\frac{\vec{I}}{m}+eB\vec{p}\times\vec{q}$\\
\hline
magnetic~fields&$e^*B^*=\frac{1}{m\omega}~,~\vec{B}^*=B^*\vec{e}_3$&$eB=m\omega~,~\vec{B}=B\vec{e}_3$\\
&$\vec{A}^*=\frac{1}{2}\vec{B}^*\times \vec{p}$~(\textbf{potential})&$\vec{A}=\frac{1}{2}\vec{B}\times \vec{q}$~(\textbf{potential})\\
\hline
noncommutative &$(s,p_i,\alpha,q^i)$&$(s,I_i,\alpha,q^i)$\\
phase~coordinates~&\textbf{with}~$\vec{q}=\frac{\vec{k}}{m},~\alpha=arctg(\frac{f_2}{f_1})$~&$\textbf{with}~\vec{q}=\frac{\vec{k}}{m},
~\alpha=arctg(\frac{p_2}{p_1})$~\\
~&~~~~~ $s=j+\vec{p}\times \vec{q}-e^*B^*\frac{\vec{p}^2}{2}$&~~~~~~$s=j+\vec{I}\times \vec{q}-eB\frac{\vec{q}^2}{2}$\\
\hline
symplectic~form&$\sigma_1=\sigma_0+G^{ij}dp_i\wedge dp_j$&$\sigma_1^{\prime}=\sigma_O^{\prime}+F_{ij}dq^i\wedge dq^j$\\
&\textbf{with}~$G^{ij}=e^*B^*\epsilon^{ij}$&\textbf{with}~$F_{ij}=eB\epsilon_{ij}$\\
\hline
Poisson~ brackets&$\{s,p_j\}=p_i\epsilon^i_j$~&$\{s,I_j\}=(I_i-eA_i)\epsilon^i_j~$\\
~&$\{s,q^i\}=\epsilon^i_j(q^j-e^*A^{*j})$&$\{s,q^i\}=\epsilon^i_jq^j$\\
~&$\{p_i,p_j\}=0$&$\{I_i,I_j\}=-eB\epsilon_{ij}$\\
~&$\{p_i,q^j\}=\delta^j_i$&$\{I_i,q^j\}=\delta^j_i$\\
~&$\{q^i,q^j\}=-e^*B^*\epsilon^{ij}$&$\{q^i,q^j\}=0$\\
\hline
hamiltonian function&$H=\frac{\vec{p}^{~2}}{2m}+V(\vec{q},\alpha)+e^*\vec{f}.(\vec{p}\times \vec{B}^*)$&$H=\vec{I}.\frac{\vec{p}}{m}+V(\vec{q},
\alpha)+e\vec{p}.(\vec{B}\times \vec{q})$\\
\hline
potential energy&$V(\vec{q},\alpha)=-\vec{f}.\vec{q}-\frac{K^*}{m}\alpha$&$V(\vec{q},\alpha)=\mp\frac{m\omega^2\vec{q}^2}{2}-\frac{K}{m}\alpha$\\
\hline
equations of motion&$\frac{d\alpha}{dt}=0$~,$\frac{ds}{dt}=\frac{K^*(t)}{m}$&$\frac{d\alpha}{dt}=0$,~$\frac{ds}{dt}=\frac{K(t)}{m}$\\
~&$\frac{d\vec{q}}{dt}=\frac{\vec{p}}{m}$~,$\frac{d\vec{p}}{dt}=\vec{f} (constant)~~~~~$ & $\frac{d\vec{q}}{dt}=\frac{\vec{p}}{m}$ (constant)~,
$\frac{d\vec{I}}{dt}=\pm m\omega^2\vec{q}$\\
~&\textbf{with}~&\textbf{with}\\
~&$K^*(t)=K^*+N^*t$&$~K(t)=K+Nt$\\
~&$K^*=m\vec{q}\times \vec{f}+e^*mB^*\vec{p}.\vec{f}$&$~K=\vec{I}\times \vec{p}-eB\vec{q}.\vec{p}$\\
~&~~~~~~$N^*=\vec{f}\times \vec{p}-e^*B^*m\frac{f^2}{2}$&~~~~~~~$N=\pm m\omega^2\vec{q}\times \vec{p}\mp eB\frac{\vec{p}^2}{2m}$\\
\hline
"Newton's equations"&$\frac{d^2s(t)}{dt^2}=\frac{N^*}{m}~,~\frac{d^2\vec{q}(t)}{dt^2}=\frac{\vec{f}}{m}$&$\frac{d^2s(t)}{dt^2}=\frac{N}{m}~,~
\frac{d^2\vec{I}(t)}{dt^2}=\pm\omega^2\vec{p}$\\
\hline
canonical&$(s,p_i,\alpha,\tilde{q}^i)$&$(s,\tilde{I}_i,\alpha,q^i)$\\
phase~coordinates&\textbf{with}~$\tilde{q}^i=q^i-e^*A^{*i}$&$\textbf{with}~\tilde{I}_i=I_i-eA_i$\\
\hline
\end{tabular}
\end{eqnarray*}

\end{document}